\documentclass[a4paper,11pt]{article}
\pdfoutput=1 

\usepackage{jheppub} 

\usepackage[T1]{fontenc} 

\title{\boldmath Study change of the performance of airfoil of small wind turbine under low wind speed by CFD simulation}

\author[a,b]{Le Quang Sang,}
\author[c]{Dinh Van Thin,}
\author[c,1]{Nguyen Huu Duc\note{Corresponding author.},}
\author[a]{Nguyen Duc Minh,}
\author[a]{Doan Hong Quan,}
\author[a]{and Le Thi Thuy Hang}

\affiliation[a]{Institute of Energy Science, Vietnam Academy of Science and Technology, Hanoi, Vietnam}
\affiliation[b]{Graduate University of Science and Technology, Vietnam Academy of Science and Technology, Ha Noi, Vietnam}
\affiliation[c]{Faculty of Energy Technology, Electric Power University, 235 Hoang Quoc Viet, Hanoi, Vietnam}

\emailAdd{lequangsang@ies.vast.vn}
\emailAdd{thindv@epu.edu.vn}
\emailAdd{ducnh@epu.edu.vn}
\emailAdd{ndminh@ies.vast.vn}
\emailAdd{doanhongquanepu@gmail.com}
\emailAdd{lehang@ies.vast.vn}

\abstract{Renewable energy has received strong attention and investment to replace fossil energy sources and reduce greenhouse gas emissions. Quite good and good wind speed areas have been invested in building large-capacity wind farms for many years. The low wind speed region occupies a very large on the world, which has been interested in the exploitation of wind energy in recent years. In this study, the original airfoil of S1010 operated at low wind speed was redesigned to increase the aerodynamic efficiency of the airfoil by using XFLR5 software. After, the new VAST-EPU-S1010 airfoil model was adjusted to the maximum thickness and the maximum thickness position. It was simulated in low wind speed conditions of 4-6 m/s by CFD simulation. The lift coefficient, drag coefficient and $C_{L}$/$C_{D}$ coefficient ratio were evaluated under the effect of the angle of attack and the maximum thickness by using the $k-\epsilon$ model. Simulation results show that the VAST-EPU-S1010 airfoil achieved the greatest aerodynamic efficiency at the angle of attack of $3\,^{\circ}$, the maximum thickness of 8\% and the maximum thickness position of 20.32\%. The maximum value of $C_{L}$/$C_{D}$ of the new airfoil at 6 m/s is higher than at the 4 m/s by about 6.25\%.}

\begin{document} 
\maketitle
\flushbottom

\section{Introduction}

Wind energy is one of the most developed renewable energy sources. Areas with medium and high wind speeds have exploited in most of the convenient locations. Specifically, according to the report of the GWEC organization, the onshore installed wind power capacity is about 68.8 GW, and the offshore wind power capacity has gone beyond 8.8 GW in 2022~\cite{d1}. This growth increases by 9\% compared with the 2021 year. 

However, the areas of low wind speed with a very large area in the world have not been exploited yet. Therefore, to effectively exploit wind energy in low wind speed areas, more research is needed to come up with reasonable and high-efficiency wind turbine blades. Research results on wind turbines in the low wind speed region at the NREL have shown several criteria such as longer blades, taller towers, and modern control methods~\cite{d2}. The purpose of designing this wind turbine technology is to increase the output of electricity produced and reduce the investment rate. This also leads to an increase in blade mass and affects the aerodynamic efficiency of the entire wind turbine. 

The IEC61400-1 standard is divided into 4 wind class categories and corresponding to 4 different wind turbine technologies~\cite{d3}. Type 1 is suitable for areas with high wind speed above 10 m/s, type 2 is suitable for average wind speed of 8.5 m/s or more, type 3 is suitable for low wind speed of 7.5 m/s or more, and category 4 is according to actual requirements. In previous studies, the high-power wind turbine blades in the low wind speed region are often designed based on the wind turbine blades in the high wind speed region~\cite{d4,d5,d6}.

To fully design a wind turbine blade, aerodynamic design and blade load design are required. Aerodynamic design is influenced by the outer shape of the wind turbine blade, most aerodynamic design goals are to maximize the ability to harness energy in the wind to maximize the power output of the wind turbine. wind turbines. Airfoil load design depends on the internal structure of the airfoil and takes into account characteristics such as durability, stable operation, and vibration~\cite{d7,d8}. There are many methods of designing and simulating wind turbine blades such as finite element method (FEM), FAST software, Qblade… Specifically, to reduce wind turbine blade weight and analyze blade load, the authors have using a combination of FAST software and particle swarm optimization (PSO) algorithm~\cite{d9}. Another study used BEM, FEM theory and multi-target genetic algorithm to get maximum power output and minimum blade mass~\cite{d10}. The aerodynamic design parameters and wind turbine blade structure are considered variables in the BEM and FEM combination algorithm~\cite{d11}. For small wind turbines, the starting wind speed and power output are the objective functions for simulation. Chord and torsion angle are considered as variables to maximize the power of small wind turbines~\cite{d12}.

The field of designing wind turbine blades to work well in low wind speed conditions has also been interesting for many years. Many organizations and wind energy research centers in the world have launched some original blade models operating in low wind speed areas but for specific applications of those airfoils in different areas. more research is needed. Specifically, NREL introduced 7 blade pattern sequences with 23 different models for horizontal axis wind turbines~\cite{d13}. Delft University of Technology (DUT) wind turbine blade models have chord thicknesses ranging from 15-40\%~\cite{d14}. Three series of Risø blade samples are presented and tested in wind tunnels for wind turbines of several hundred kilowatts to megawatts~\cite{d15}.

Besides, for small wind turbines, the Re number is usually less than  $5\times10^{5}$~\cite{d16,d17}. Blade models operating in the low wind speed region will often have a flow separation from the wind turbine blade's upper surface. This flow separation is due to an adverse pressure differential across the blade surface by the influence of the blade thickness. The vortex generator mounted on the surface of the wind turbine blade has tapered and eliminated the delamination bubble at the higher angle of attack~\cite{d18}. The thickness of wind turbine blades in low wind speed conditions is designed to be thinner to reduce adverse pressure differentials~\cite{d19,d20}. Research on airfoils of the form SG6040 – SG6043 at Re coefficient from $1\times10^{5}$ to $5\times10^{5}$, showed good performance~\cite{d21}. On the other hand, some optimization studies to calculate the shape of wind turbine blades focus on increasing the radius of the tip and the curvature of the wind turbine blades to reduce~\cite{d22,d23,d24}. An important issue with wind turbine blades operating at low wind speeds is having a good starting torque~\cite{d25,d6}. Thus, to optimize the aerodynamics and structure of the wind turbine blade model at low wind speeds, it is necessary to consider parameters such as chord, torsion, tip radius, and thickness.

Many studies have shown the effect of blade thickness on aerodynamic performance. A sample of 18\% thicker wind turbine blades operating in the Re coefficient region from $2\times10^{5}$ to $3\times10^{5}$ showed higher efficiency than the original blade model in low wind speed conditions~\cite{d26}. Another study showed that when the maximum thickness was increased to 28.5\%, the power factor did not change, and the starting torque was better (i.e, the starting wind speed decreased)~\cite{d27,d28}. However, when the thickness of the wind turbine blades exceeds the permissible limit, it will cause difficult starting torque due to the large drag coefficient, particularly in the low AOA region. On the other hand, The NACA 2412 airfoil works well in low wind speed conditions with a maximum camber of 0.02 times and maximum thickness of 0.12 times chord~\cite{d29,d30}. Another research of the NACA 2412 showed that the lift increased at various angles of attack~\cite{d31}.

The computational fluid dynamics (CFD) simulation has been widely used to analyze the influence of airflow on the blade model and the entire blade of a wind turbine. This method can accurately simulate boundary layers on the whole airfoil. The previous studies that evaluated the effects of thickness and curvature on the aerodynamic performance of airfoils are still limited at low wind speeds. 

Therefore, this study will change the shape of the airfoil. Then, the new airfoil is simulated in the low wind speed region to estimate the aerodynamic performance of the airfoil. The publicized original airfoils was considered and selected from famous organizations\cite{d13,d14,d15,d32,d33}. The focused parameters for improving aerodynamic performance are maximum thickness, maximum thickness position, maximum curvature, and maximum curvature position. The results of this study help to better understand the new airfoil design method, and improve the blade performance in low wind speed. From there, it is possible to build a complete wind turbine blade with high aerodynamic efficiency for further studies.


\section{Materials and Methods}

This section describes a method for designing a wind turbine blade model using the XFLR5 and CFD simulation tools. The blade models were selected from the original blade models that operate stably in low wind speed conditions. The parameters of maximum thickness and maximum thickness position are analyzed to obtain a better $C_{L}$/$C_{D}$ ratio. The low wind speed is from 4 m/s to 6 m/s.

\subsection{The method of selecting and designing the airfoil}
\begin{figure}[!htbp]
\centering 
\includegraphics[width=0.45\textwidth]{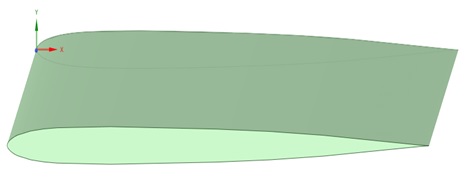}
\caption{\label{fig:01}3D design of S1010 airfoil}
\end{figure}

Many previous researches have published suitable wind turbine blades in low wind speed conditions. A S1010 airfoil was selected for this research as shown in Figure~\ref{fig:01}. 

The criteria for selecting blade models suitable for the low wind speed conditions as follow:

\begin{itemize}
\item Selection of airfoil models that work well in low wind speed conditions from previous studies. Get data on original airfoil samples from websites~\cite{d34};
\item Import original data into XFLR5 software, using reverse engineering method to change parameters of the maximum thickness and the maximum thickness position;
\item New airfoil model is built and simulated by CFD simulation using the $k-\epsilon$ model for analysis, and the results are compared with the original airfoil data;
\item $C_{L}$, $C_{D}$ coefficient, and $C_{L}$/$C_{D}$ ratio are calculated in low wind speed conditions from 4 m/s to 6m/s;
\item The size of the airfoil is designed with dimensions: Chord length of 304.8 mm and Span of 850 mm;
\end{itemize}

\subsection{XFLR5 software}
XFLR5 software is used to analyze wind turbine blade at low Re number. Lift coefficient, drag coefficient, and lift/thrust coefficient ratio are evaluated with different angles of attacks at low Re number.

When the number of panels has been determined, XFLR5 will calculate the velocity and the AoA values on the surface of each panel, thereby determining the values of vortex strength ($\gamma$), lift force ($L$) and drag force ($D$) over the entire surface of the airfoil. Characteristic values such as pressure coefficient ($C_{p}$), lift coefficient ($C_{L}$), drag coefficient ($C_{D}$) of the airfoil are determined:

\begin{equation}
\label{eq:01}
C_{L} = \frac{2L}{\rho \, V_{\infty}^2 \, c}
\end{equation}

\begin{equation}
\label{eq:02}
C_{D} = \frac{2D}{\rho \, V_{\infty}^2 \, c}
\end{equation}

\begin{equation}
\label{eq:03}
C_{p} = 1-\left(\frac{\gamma}{V_{\infty}}\right)^2
\end{equation}

Where: $c$ is a chord of the airfoil [$m$]; $\rho$ is the density of air [$kg/m^3$]; $V_{\infty}$ is the speed of free flow [$m/s$].

The formula for calculating Reynolds number to the flow velocity at any position in the model is given as follows:

\begin{equation}
\label{eq:04}
Re = \frac{\rho \, v \, c}{\mu}
\end{equation}

\subsection{CFD software}
In this study, CFD simulation software was used to simulate the split flows on the surface of the airfoil.

\begin{figure}[!htbp]
\centering 
\includegraphics[width=0.45\textwidth]{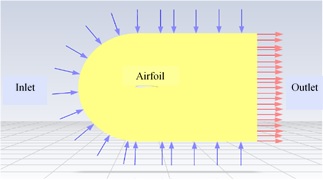}
\caption{\label{fig:02}Computational domain around VAST-EPU-S1010 airfoil}
\end{figure}

The airfoil was built in the two-dimensional model using SpaceClaim graphics software. The distance from the inlet layer to the tip of the airfoil is 6 m, and the distance from the endpoint of the airfoil to the outlet layer is 18 m. Boundary conditions include the inlet surface, the outlet surface, the airfoil's surface, and the model's symmetry surface as shown in the Figure~\ref{fig:02}.

\begin{figure}[!htbp]
\centering 
\includegraphics[width=0.45\textwidth]{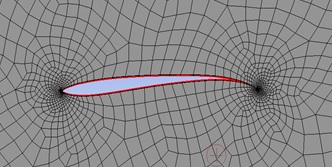}
\caption{\label{fig:03}Mesh around VAST-EPU-S1010 airfoil}
\end{figure}

\begin{figure}[!htbp]
\centering 
\includegraphics[width=0.45\textwidth]{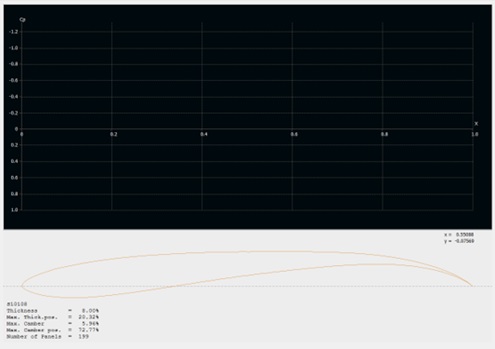}
\caption{\label{fig:04}Input data of the VAST-EPU-S1010 airfoil}
\end{figure}

The models are meshed with a combination of triangular and quadrangular grids, with an average mesh size of 0.03 m. The boundary condition layer surrounding the airfoil is divided into a quadrangular grid, the size of the first mesh layer adjacent to the surface of the airfoil is 0.005 m. The total number of these boundary condition classes is five layers, and each layer will have a difference of 1.1 times as exhibited in Figure~\ref{fig:03} and Figure~\ref{fig:04}. 

The total number of meshes of the models built in this paper is approximately 300,000 meshes, and the mesh quality of each model is 0.95 or more. This value ensures the convergence of the models. The turbulence model used in this study is the Realizable $k-\epsilon$ Turbulence Model, see~\eqref{eq:05}.

\begin{equation}
\label{eq:05}
\frac{\partial (\rho \epsilon)}{\partial t} + \frac{\partial (\rho \epsilon u_{j})}{\partial x_{j}}=\frac{\partial}{\partial x_{j}} \left[ \left( \mu + \frac{\mu _{t}}{\sigma _{\epsilon}} \right) \frac{\partial \epsilon}{\partial x_{j}} \right] + \rho \, C_{1} S \, \epsilon - C_{2} \rho \frac{\epsilon ^2}{k + \sqrt{v \epsilon}} + C_{1 \epsilon} \frac{\epsilon}{k} C_{3 \epsilon} G_{b}
\end{equation}

Where: $C_{1} = max \left[ 0.43 \frac{\eta}{\eta +5} \right]$ and $\eta = S \frac{k}{\epsilon} $; S is the scalar measure of the deformation. The values $\sigma _{k}$, $\sigma _{\epsilon}$, $C_{1 \epsilon}$, $C_{3 \epsilon}$, $C_{1}$ and $C_{2}$ are constants.

\section{Results and discussion}

\begin{figure}[!htbp]
\centering 
\includegraphics[width=0.45\textwidth]{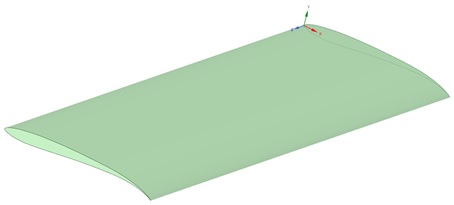}
\caption{\label{fig:05}3D design of VAST-EPU-S1010 airfoil}
\end{figure}

After using XFLR5 to redesign from the original airfoil of S1010, a new airfoil model of VAST-EPU-S1010 is proposed as shown in Figure~\ref{fig:05}. The parameters of the airfoil of S1010 and VAST-EPU-S1010 are listed in Table~\ref{tab:01}. 

\begin{table}[!htbp]
\centering
\begin{tabular}{|l|c|c|}
\hline
  & S1010 & VAST-EPU-S1010\\
\hline 
Thickness (\%) & 6.02 & 8.00 \\
Max. Thick. Pos. (\%) & 23.42 & 20.32 \\
Max. Camber (\%) & 0.00 & 5.96 \\
Max. Cam. Pos. (\%) & 0.50 & 72.77 \\
\hline
\end{tabular}
\caption{\label{tab:01}Airfoil model parameters of S1010 and VAST-EPU-S1010}
\end{table}

The new airfoil has the maximum thickness, maximum thickness position, maximum camber and maximum camber position are changed. Simulation results of the aerodynamic performance of the new airfoil are performed.

\subsection{Low wind speed region}
\begin{figure}[!htbp]
\centering 
\includegraphics[width=0.7\textwidth]{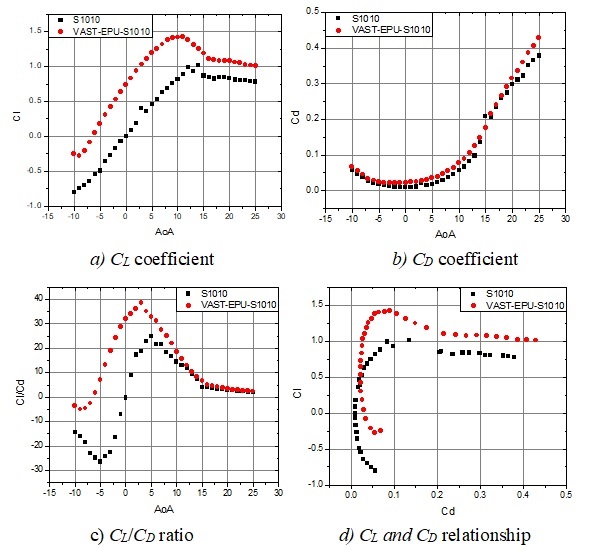}
\caption{\label{fig:06}Effect of AoA on $C_{L}$, $C_{D}$ coefficients and $C_{L}/C_{D}$ ratio at wind speed of 4 m/s}
\end{figure}

\begin{figure}[!htbp]
\centering 
\includegraphics[width=0.7\textwidth]{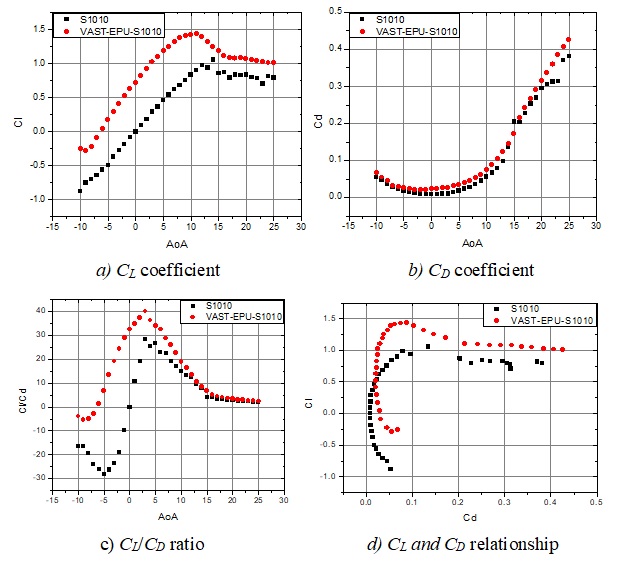}
\caption{\label{fig:07}Effect of AoA on $C_{L}$, $C_{D}$ coefficients and $C_{L}/C_{D}$ ratio at wind speed of 5 m/s}
\end{figure}

\begin{figure}[!htbp]
\centering 
\includegraphics[width=0.7\textwidth]{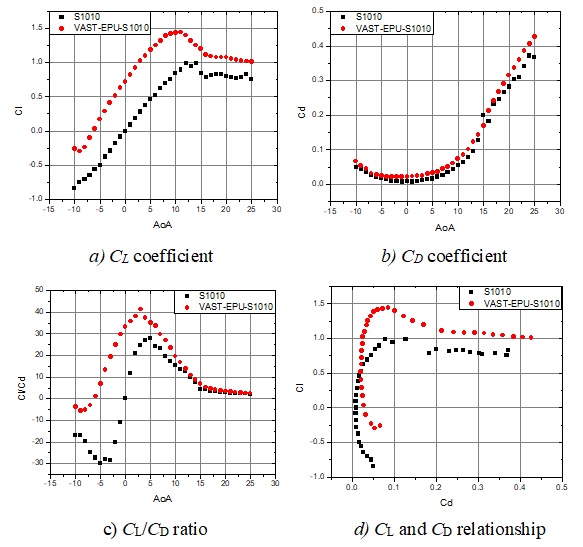}
\caption{\label{fig:08}Effect of AoA on $C_{L}$, $C_{D}$ coefficients and $C_{L}/C_{D}$ ratio at wind speed of 6 m/s}
\end{figure}

In this section, the effect of wind speed on aerodynamic performance is evaluated. The lift coefficient, drag coefficient, and lift/drag coefficient ratio are considered when the AoA varies from $-10\,^{\circ}$ to $15\,^{\circ}$. The analysis parameters of Ansys Fluent software for two airfoils are the same, only changing the wind speed from 4-6 m/s. 
From Figures~\ref{fig:06} to~\ref{fig:08}, graphs of a and b show the lift and drag coefficients when the AoA changes, and graph c examines the effect of AoA on the lift/drag coefficient ratio. Finally, graph d shows the relationship between the lift and drag coefficient. 

\begin{figure}[!htbp]
\centering 
\includegraphics[width=0.45\textwidth]{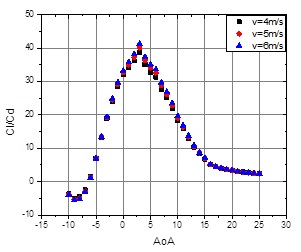}
\caption{\label{fig:09}Effect of the wind speed on the maximum $C_{L}/C_{D}$ ratio values}
\end{figure}

For graph $a$ of Figures~\ref{fig:06} to~\ref{fig:08}, the lift coefficient, $C_{L}$ reached the maximum values of 1.422, 1.433, and 1.442 with wind speeds of 4-6 m/s at the angles of attack of $11\,^{\circ}$, respectively. The tendency of the lift coefficient to increase rapidly from $-9\,^{\circ}$ to $11\,^{\circ}$, then decrease in the range of $11\,^{\circ}$ to $15\,^{\circ}$, and finally decrease slightly at the remaining angles of attack. The simulation results show that the maximum lift coefficient value for the new airfoil of VAST-EPU-S1010 is 35.7\% - 45.5\% higher than that of the original airfoil of S1010.

For graph $b$, the value of the drag coefficient $C_{D}$ decreases gradually in the range of AoA from $-10\,^{\circ}$ to $0\,^{\circ}$, increases slightly for the angle of attack from $0\,^{\circ}$ to $5\,^{\circ}$, and finally increases rapidly at the higher angles of attack. Graph $d$ shows the relationship between the of $C_{L}$ and $C_{D}$ coefficients in the simulation results, the new airfoil has a higher value than the original airfoil. Next, to consider the optimal value of the aerodynamic efficiency, graph $c$ represents the maximum $C_{L}/C_{D}$ ratio values of 38.74, 40.14, and 41.32 at the AoA of $3\,^{\circ}$ with the wind speeds of 4, 5, 6 m/s, respectively, as shown in Figure~\ref{fig:09}.

\subsection{Effect of angle AoA}
\begin{figure}[!htbp]
\centering 
\includegraphics[width=0.8\textwidth]{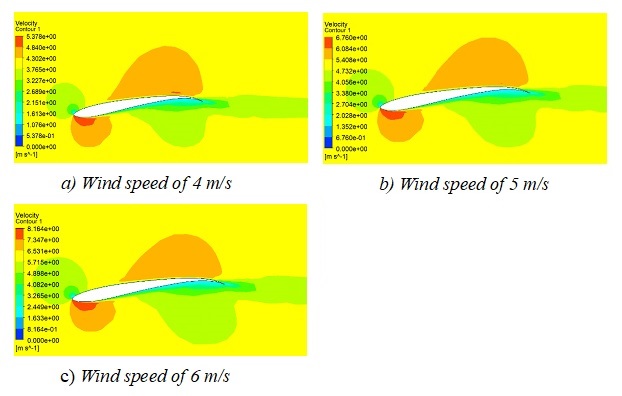}
\caption{\label{fig:10}Effect of wind speed on velocity contour at AoA = $-5\,^{\circ}$}
\end{figure}

\begin{figure}[!htbp]
\centering 
\includegraphics[width=0.8\textwidth]{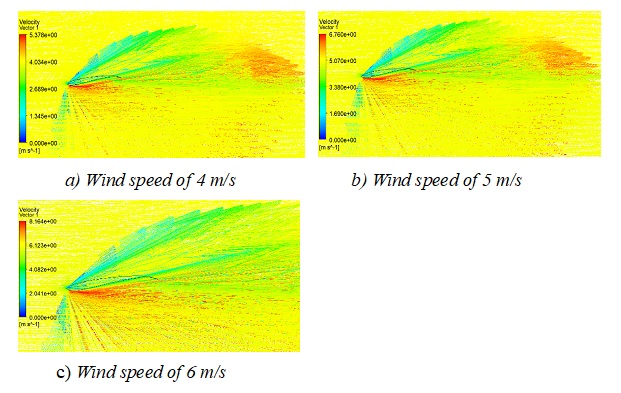}
\caption{\label{fig:11}Effect of wind speed on velocity vector at AoA = $-5\,^{\circ}$}
\end{figure}

\begin{figure}[!htbp]
\centering 
\includegraphics[width=0.8\textwidth]{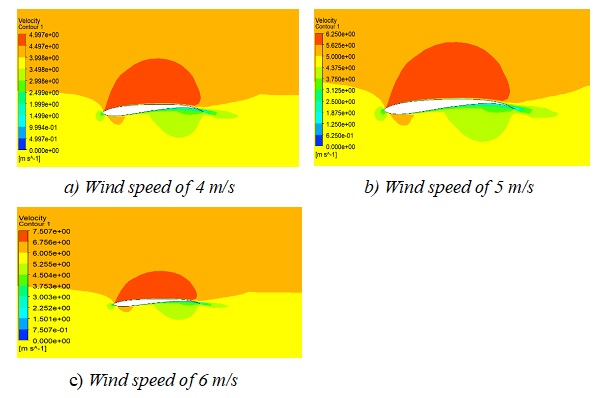}
\caption{\label{fig:12}Effect of wind speed on velocity contour at AoA = $0\,^{\circ}$}
\end{figure}

\begin{figure}[!htbp]
\centering 
\includegraphics[width=0.8\textwidth]{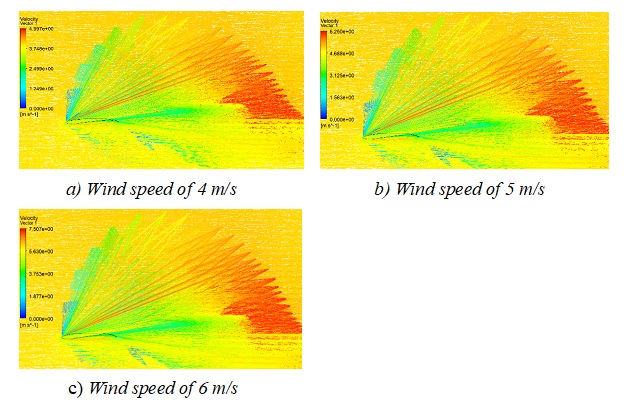}
\caption{\label{fig:13}Effect of wind speed on velocity vector at AoA = $0\,^{\circ}$}
\end{figure}

\begin{figure}[!htbp]
\centering 
\includegraphics[width=0.8\textwidth]{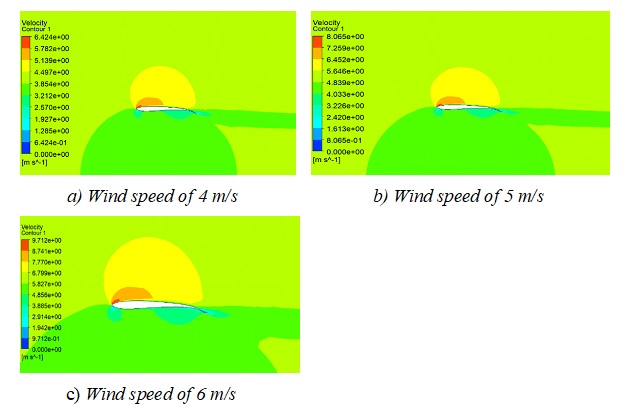}
\caption{\label{fig:14}Effect of wind speed on velocity contour at AoA = $5\,^{\circ}$}
\end{figure}

\begin{figure}[!htbp]
\centering 
\includegraphics[width=0.8\textwidth]{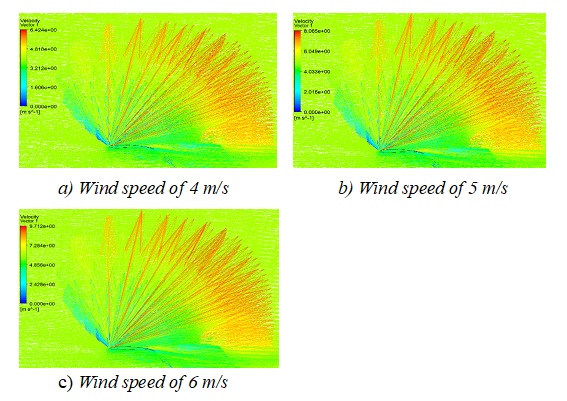}
\caption{\label{fig:15}Effect of wind speed on velocity vector at AoA = $5\,^{\circ}$}
\end{figure}

\begin{figure}[!htbp]
\centering 
\includegraphics[width=0.8\textwidth]{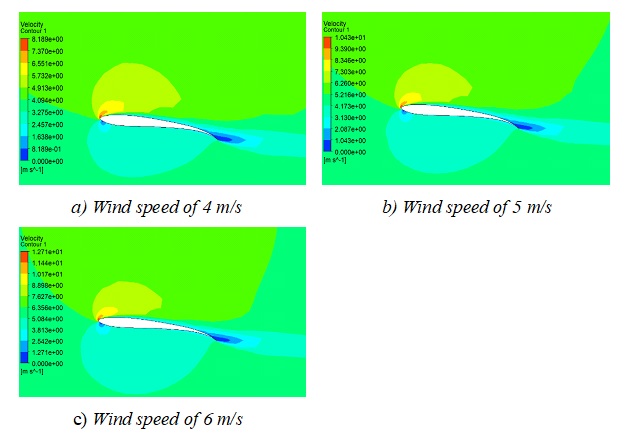}
\caption{\label{fig:16}Effect of wind speed on velocity contour at AoA = $10\,^{\circ}$}
\end{figure}

\begin{figure}[!htbp]
\centering 
\includegraphics[width=0.8\textwidth]{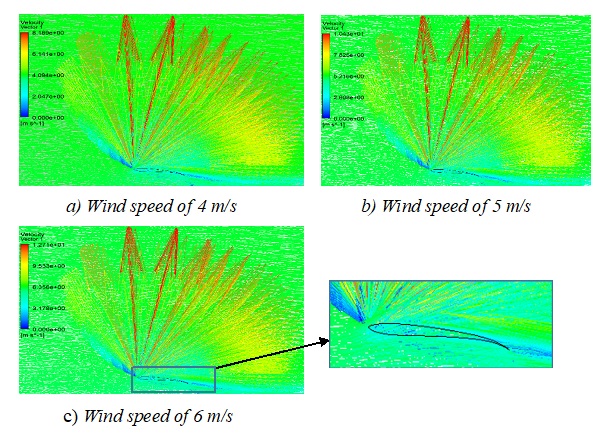}
\caption{\label{fig:17}Effect of wind speed on velocity vector at AoA = $10\,^{\circ}$}
\end{figure}

\begin{figure}[!htbp]
\centering 
\includegraphics[width=0.8\textwidth]{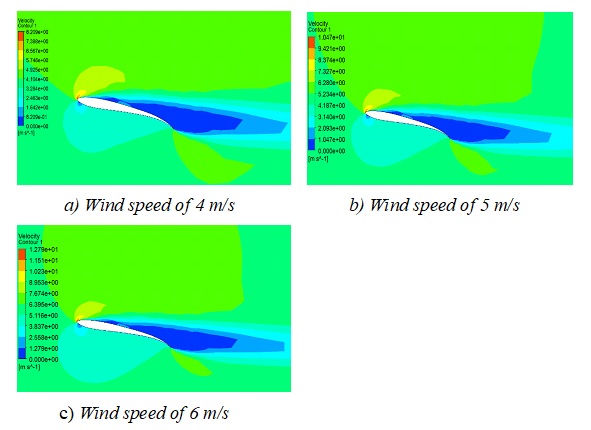}
\caption{\label{fig:18}Effect of wind speed on velocity contour at AoA = $15\,^{\circ}$}
\end{figure}

To understand more about the effect of the AoA on the performance of the airfoil, the velocity contours and velocity vectors around the airfoil will be evaluated. Simulation data considered at angles of attack from $-5\,^{\circ}$ to $15\,^{\circ}$ with steps of $5\,^{\circ}$ are shown in Figure~\ref{fig:10} to Figure~\ref{fig:18}.

\begin{figure}[!htbp]
\centering 
\includegraphics[width=0.8\textwidth]{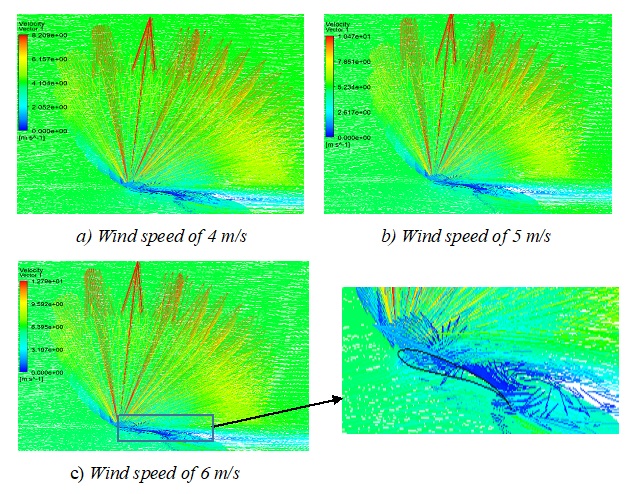}
\caption{\label{fig:19}Effect of wind speed on velocity vector at AoA = $15\,^{\circ}$}
\end{figure}

The velocity contours are shown in Figures~\ref{fig:10},~\ref{fig:12},~\ref{fig:14},~\ref{fig:16}, and~\ref{fig:18} for the angles of attack of $-5\,^{\circ}$, $0\,^{\circ}$, $5\,^{\circ}$, $10\,^{\circ}$, and $15\,^{\circ}$, respectively. In Figure~\ref{fig:10}, the airfoil edge underside and the tail upperside have a high velocity which is indicated by the orange color. In Figure~\ref{fig:12}, around the maximum $C_{L}/C_{D}$ ratio value, the red and orange colors can be seen covering the entire upper surface of the airfoil at the angle of attack of $0\,^{\circ}$, which means that the aerodynamic efficiency is great within this AoA region. Figure~\ref{fig:14} shows that around the upper surface of the airfoil, the yellow color covers most of the surface, the orange color occupies about 40\% of the upper surface area, and the red color remains very little at the tip of the airfoil. Figure~\ref{fig:16} exhibits the beginning of the zero-velocity region on the underside of the edge of the airfoil and the tail of the upper of the airfoil. At an angle of attack of $10\,^{\circ}$, a delamination layer began to appear on the surface of the airfoil at the tail. Specifically, the velocity direction of this region will be evaluated by the velocity vectors in the next section. Figure~\ref{fig:18} shows the area of flow velocity equal to 0 extending to almost the entire upper surface of the airfoil. This shows that the drag coefficient has increased significantly in the high angle of attack region.

The velocity vectors are shown in Figures~\ref{fig:11},~\ref{fig:13},~\ref{fig:15},~\ref{fig:17}, and~\ref{fig:19} for different angles of attack, respectively. The direction of the vector and the magnitude of the velocity are shown in colors from blue to red. At $0\,^{\circ}$ angle of attack, the red and orange velocity vectors account for a large proportion, at other angles of attack, this number of vectors is significantly reduced as the angle of attack is higher. At the $10\,^{\circ}$ angle of attack, velocity vectors with different directions appeared at the tail end of the airfoil. Seen more clearly with a $15\,^{\circ}$ angle of attack, the magnitude, and direction of the velocity vector form vortices on the surface of the airfoil. This phenomenon occurs at a high angle of attack because the friction force of the airflow with the surface of the airfoil increases significantly against the inertia force of the wind.

With the above simulation results, it can be seen that when increasing the larger angle AoA, the delamination layer begins to appear gradually from the tail of the airfoil. The direction and magnitude of the eddies are expressed as the velocity vectors. The higher the angle of attack, the more turbulent the eddies on the surface of the airfoil. This phenomenon will cause significant drag on the airfoil.

\subsection{Evaluation of the effect of thickness}
This section indicates the effect of thickness on the airfoil performance. The $C_{L}/C_{D}$ ratio is evaluated at the optimal angle of attack of $3\,^{\circ}$ at wind speeds of 4-6 m/s. The simulation results are shown in Table~\ref{tab:02}. 

\begin{table}[!htbp]
\centering
\begin{tabular}{|c|c|c|c|}
\hline
Thickness & \multicolumn{3}{|c|} {$C_{L}/C_{D}$ ratio} \\
\cline{2-4} (\%) & at 4 m/s & at 5 m/s & at 6 m/s \\
\hline 
6 & 36.608 & 37.852 & 38.882 \\
7 & 36.746 & 37.998 & 39.035 \\
8 & 38.740 & 40.141 & 41.323 \\
9 & 35.219 & 36.159 & 37.266 \\
\hline
\end{tabular}
\caption{\label{tab:02}$C_{L}/C_{D}$ ratio value at angle AoA = $3\,^{\circ}$ according to different blade thickness}
\end{table}

\begin{figure}[!htbp]
\centering 
\includegraphics[width=0.45\textwidth]{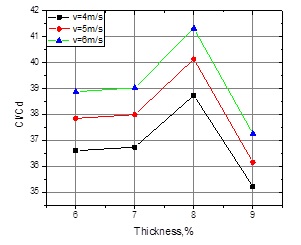}
\caption{\label{fig:20}Optimal value of $C_{L}/C_{D}$ ratio at the angle AoA = $3\,^{\circ}$ according to different maximum thickness}
\end{figure}

\begin{figure}[!htbp]
\centering 
\includegraphics[width=0.8\textwidth]{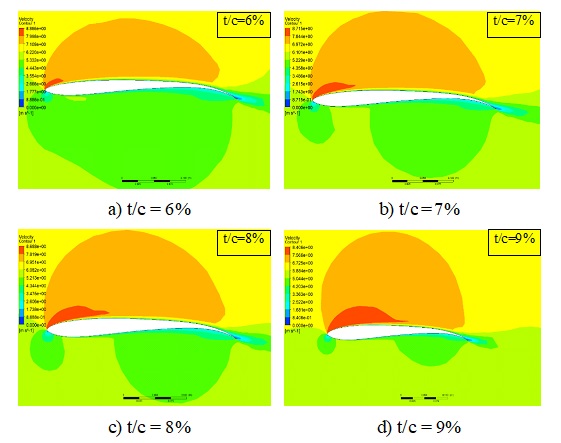}
\caption{\label{fig:21}Velocity contour with thickness varying from 6\% - 9\%}
\end{figure}

At the wind speed of 4 m/s, the maximum value of the $C_{L}/C_{D}$ ratio reached 38.739 at maximum thickness of 8\%. Similarly, at wind speeds of 5 m/s and 6 m/s, the maximum values of the $C_{L}/C_{D}$ ratio are 40,141 and 41,323 at the maximum thickness of 8\%, respectively, as shown in Figure~\ref{fig:20}.

From Figure~\ref{fig:21}, it can be seen that as the blade thickness increases, the surface velocity increases, represented by the red and orange areas. Simulated results at optimal angle of attack $3\,^{\circ}$ and the wind speed of 6 m/s. Looking at the figure, the difference in upper and lower surface pressure at the maximum thickness of 8\% is higher than other thicknesses, resulting in the highest aerodynamic efficiency of the airfoil model.

\section{Conclusions}

In this paper, the new wind turbine blade model VAST-EPU-S1010 indicated based on the original blade model of S1010. The maximum thickness and maximum thickness position were considered. The performance of the new airfoil was simulated by using CFD under the low wind speeds. Lift coefficient, drag coefficient, and lift/drag ratio have been evaluated under the influence of different angles of attack from $-10\,^{\circ}$ to $15\,^{\circ}$. Some of the results are summarized as follows:

\begin{itemize}
\item The new airfoil of VAST-EPU-S1010 has been designed and simulated in low wind speed conditions of 4-6 m/s. The maximum value of the lift/drag ratio of the new airfoil increased from 35.7\% to 45.5\% compared to the original airfoil model. The maximum value of $C_{L}/C_{D}$ is 41.32 at the wind speed of 6 m/s. Besides, the wind speed also affected to $C_{L}/C_{D}$ ratio, the maximum value of $C_{L}/C_{D}$ at 6 m/s is higher than at the 4 m/s by about 6.25\%.
\item At the low wind speed, the influence of the angle of attack from $-5\,^{\circ}$ to $15\,^{\circ}$ has also been simulated and evaluated. The results show that the airfoil efficiency around the optimal angle of attack of $3\,^{\circ}$ is the highest, the airfoil efficiency decreases as the angle of attack increases. From the angle of attack of $10\,^{\circ}$ or more, the tail of the airfoil appears as a separate layer and the velocity vector moves in different directions. The higher the angle of attack, the earlier the separation layer appears and occupies a larger surface area of the airfoil.
\item Thickness also affects the aerodynamic performance of the airfoil. The new airfoil model of VAST-EPU-S1010 is optimized at a maximum thickness of 8\% and a maximum thickness position of 20.32\%.
\end{itemize}

\paragraph{Future work:} With the above simulation results, it can be seen that there is still a lot of opportunity to improve the efficiency of the wind turbine and reduce the starting wind speed for the wind turbine in low wind speed conditions. With increasingly modern simulation tools and the application of artificial intelligence algorithms, the ability to optimize the blade pattern and the entire wind turbine blade will increase. Soon, we will develop an optimal design method for wind turbine blades to improve power production efficiency in low wind speed conditions. The method will apply a combination of artificial intelligence tools to select the optimal parameters of chord, thickness, and airfoil length. The simulation results will be compared with the original airfoil and the complete design of wind turbine blade using the new airfoil.

\acknowledgments

This work was supported by the Institute of Energy Science (IES) which belongs to Vietnam Academy of Science and Technology (VAST). This research was funded by Vietnam Academy of Science and Technology (VAST), Code VAST07.01/22-23.



\begin{thebibliography}{99}

\bibitem{d1}
GWEC, \emph{GLOBAL WIND REPORT 2023}, The Global Wind Energy Council (2023)

\bibitem{d2}
Cohen J., Schweizer T., Laxson A., et al. 2008, \emph{Technology Improvement Opportunities for Low Wind Speed Turbines and Implications for Cost of Energy Reduction}, National Renewable Energy Laboratory, Golden, Colorado (US). Technical Report NREL/TP-500-41036 (2008)

\bibitem{d3}
IEC, \emph{Wind turbines – Part 1: Design requirements. IEC 61400-1. Version 3}, 3.	International Electrotechnical Commission (2005)

\bibitem{d4}
Singh Ronit K, Rafiuddin Ahmed M., \emph{Blade design and performance testing of a small wind turbine rotor for low wind speed applications}, \emph{Renewable Energy} {\bf 50} (2013) pg.812–9

\bibitem{d5}
Clifton-Smith M, Wood D, Wright A., \emph{Optimising wind turbine design for operation in low wind speed environments}, \emph{Wind Energy Syst} {\bf 13} (2011) pg.366–87

\bibitem{d6}
Wright AK, Wood DH., \emph{The starting and low wind speed behavior of a small horizontal axis wind turbine}, \emph{J. Wind Eng Ind Aerodyn} {\bf 92} (2004) pg.1265–79

\bibitem{d7}
Xudong W, Shen W-Z, Zhu W-J, Sørensen J-N, Jin C., \emph{Shape optimization of wind turbine blades}, \emph{Wind Energy} {\bf 12} (2009) pg.781-803

\bibitem{d8}
Maalawi KY, Negm HM., \emph{Optimal frequency design of wind turbine blades}, \emph{J Wind Eng Ind Aerodyn} {\bf 90} (2002) pg.961–86

\bibitem{d9}
Liao CC, Zhao XL, Xu JZ., \emph{Blade layers optimization of wind turbines using FAST and improved PSO}, \emph{Renewable Energy} {\bf 42} (2012) pg.227–33

\bibitem{d10}
Zhu J, Cai X, Gu R., \emph{Aerodynamic and structural integrated optimization design of horizontal-axis wind turbine blades}, \emph{Energies} {\bf 9} (2016) pg.66

\bibitem{d11}
Dal Monte Andrea, De Betta Stefano, Castelli Marco Raciti., \emph{Proposal for a coupled aerodynamic–structural wind turbine blade optimization}, \emph{Compos Struct.} {\bf 159} (2017) pg.144–56

\bibitem{d12}
Pourrajabian, A., Ebrahimi, R., Mirzaei, M., \emph{Applying micro scales of horizontal axis wind turbines for operation in low wind speed regions}, \emph{J. Energy Convers. Manag.} {\bf 87} (2014) pg.119–127

\bibitem{d13}
Tangler J.L., Somers D.M., \emph{NREL Airfoil Families for HAWTs}, National Renewable Energy Lab., Golden, CO (1995)

\bibitem{d14}
Timmer W.A., Van Rooij R., \emph{Summary of the Delft University wind turbine dedicated airfoils}, \emph{J. Sol. Energy Eng.} {\bf 125 (4)} (2003) pg.488-496

\bibitem{d15}
Fuglsang P., Bak C., Gaunaa M., et al., \emph{Design and verification of the Risø-B1 airfoil family for wind turbines}, \emph{J. Sol. Energy Eng.} {\bf 126 (4)} (2004) pg.1002 - 1010

\bibitem{d16}
Miley SJ., \emph{A catalog of low reynolds number airfoil data for wind turbine applications}, Texas: Department of Aerospace Engineering Texas AM University College Station (1982)

\bibitem{d17}
Elizondo J., Martinez J., and Probst O., \emph{Experimental study of a small wind turbine for low- and medium-wind regimes}, \emph{International Journal of Energy Research} {\bf 33} (2009) pg.309 - 26

\bibitem{d18}
Clausen PD and Wood DH., \emph{Low-Reynolds-number airfoils}, \emph{Annual Reviews of Fluid Mechanics} {\bf 15} (1983) pg.223 - 39

\bibitem{d19}
Lissaman PBS., \emph{Research and development issues for small wind turbines}, \emph{Renewable Energy} {\bf 16} (1999) pg.922-7

\bibitem{d20}
Giguere P. and Selig MS., \emph{Low reynolds number airfoils for small horizontal axis wind turbines}, \emph{Wind Engineering} {\bf 21} (1997) pg.379

\bibitem{d21}
Giguere P. and Selig MS., \emph{New airfoils for small horizontal axis wind turbines}, \emph{ASME Journal of Solar Energy Engineering} {\bf 120} (1998) pg.108 - 14

\bibitem{d22}
Anderson JD., \emph{Fundamentals of aerodynamics}, 3rd ed. New York:McGraw Hill (2001)

\bibitem{d23}
McGranahan BD. and Selig MS., \emph{Aerodynamic tests of six airfoils for use on small wind turbines}, Illinois: University of Illinois at Urbana-Champaign Urbana (2004)

\bibitem{d24}
Henriques JCC., Silva M., Estanqueiro AI., Gato LMC., \emph{Design of a new urban wind turbine airfoil using a pressure-load inverse method}, \emph{Renewable Energy} {\bf 34} (2009) pg.2728-34

\bibitem{d25}
Ozgener O. and Ozgener L., \emph{Exergy and reliability analysis of wind turbine systems: a case study}, \emph{Renewable and Sustainable Energy Reviews} {\bf 11} (2007) pg.1811 - 26

\bibitem{d26}
Healy J., \emph{The influence of blade camber on the output of vertical-axis wind turbines}, \emph{Wind Eng.} {\bf 2} (1978) pg.146–155

\bibitem{d27}
Batista N., Melício R., Marias J., and Catalão J., \emph{Self-start evaluation in lift-type vertical axis wind turbines: Methodology and computational tool applied to asymmetrical airfoils}, \emph{In Proceedings of the Power Engineering, Energy and Electrical Drives (POWERENG), Malaga, Spain} {\bf 11–13} (2011) pg.1–6

\bibitem{d28}
Alam M. and Iqbal M., \emph{A low cut-in speed marine current turbine}, \emph{J. Ocean Technol.} {\bf 5} (2010) pg.49–61

\bibitem{d29}
Sogukpinar H., \emph{The effects of NACA 0012 airfoil modification on aerodynamic performance improvement and obtaining high lift coefficient and post-stallairfoil}, \emph{In AIP conference proceedings} {\bf 1935(1)} (2018) pg.020001

\bibitem{d30}
Saxena E.S. and Kumar M.R., \emph{Design of NACA 2412 and its Analysis at Different Angle of Attacks, Reynolds Numbers, and a wind tunnel test}, \emph{International Journal of Engineering Research and General Science} {\bf 3(2)} (2015) pg.193-200

\bibitem{d31}
Venkatesan S.P., Kumar V.P., Kumar M.S. and Kumar S., \emph{Computational analysis of aerodynamic characteristics of dimple airfoil NACA 2412 at various angles of attack}, \emph{Idea} {\bf 46} (2015) pg.10

\bibitem{d32}
Grasso F., \emph{ECN Airfoils for Large Offshore Wind Turbines Design and Wind Tunnel Testing}, \emph{Wind Energy} {\bf 2016} (2017) pg.2015

\bibitem{d33}
Li X., Yang K., Bai J., et al., \emph{A method to evaluate the overall performance of the CAS-W1 airfoils for wind turbines}, \emph{J. Renew. Sustain. Energy} {\bf 5 (6)} (2013) pg.063118

\bibitem{d34}
Airfoil Tools, \emph{http://airfoiltools.com/airfoil/details?airfoil=s1010-il}, (accessed on 15th August 2023)





\end{thebibliography}
\end{document}